# XMCD under pressure at the Fe K edge on the energy dispersive beamline of the ESRF

O. Mathon[a], F. Baudelet[b,c], J.-P. Itié[b], S. Pasternak[a], A. Polian[b] and S. Pascarelli[a]

[a] European Synchrotron Radiation Facility, BP 220, 38043 Grenoble cedex, France
[b] Physique des Milieux Condensés, CNRS-UMR 7602 Université Paris VI, B 77, 75252 Paris CEDEX 05, France,
[c] SOLEIL, L'orme des Merisiers, Saint Aubin B.P. 48, 91192 Gif-sur-Yvette cedex, France.
E-mail : mathon@esrf.fr

The present paper demonstrates the feasibility of X-ray Absorption Spectroscopy (XAS) and X-ray Magnetic Circular Dichroism (XMCD) at high pressure at the Fe-K edge on the ID24 energy dispersive beamline of the ESRF.

In 3d transition metals, performing experiments at the hard X-ray K-edge rather than at the magnetically interesting soft X-ray L-edges represents, the only way to access the high pressure regime obtainable with Diamond Anvil Cells.

The simultaneous availability of a local structure (XAS) and of a magnetic (XMCD) probe on the sample in identical thermodynamical conditions is essential to study correlations between local structural and magnetic properties.

We briefly summarize the state of the art theoretical understanding of K-edge XMCD data, then illustrate the setup of beamline ID24 for high pressure XMCD experiments and underline the conditions required to perform measurements at the K-edges of 3d transition metals. Finally, we present two examples of recent high pressure results at the Fe-K edge in pure Fe and $Fe_3O_4$ powder.

## 1. Introduction

XMCD is the difference between X-ray absorption spectra obtained with right handed and left handed circular polarization respectively. For a finite XMCD signal to be measured in specific temperature-pressure conditions, the sample must present a net ferromagnetic or ferrimagnetic moment. The XMCD signal is then directly proportional to the magnetic moment on the absorber atom. Thanks to its element and orbital selectivity, XMCD has been widely used as a very useful probe of magnetic properties (for a recent review *e.g.* Stöhr (1999) or *Magnetism and Synchrotron Radiation*, 2001). In the hard X-ray regime, thanks to the higher penetration depth of the photons, it has the additional advantages of being bulk sensitive and applicable to systems subject to high pressure in Diamond Anvil Cells. In addition to the sensitivity to the magnetic moment of the dichroic signal, the associated average XAS spectrum contains information on the electronic and local structure. Thus, an XMCD experiment gives access to simultaneous information on the magnetic moment around the absorber atom, as well as its local structural and electronic properties i.e. on the sample with the same thermodynamical conditions. This is very important in the high-pressure domain where non hydrostatic components of the stresses can not be reproduced. These experiments open therefore the way to studies of correlations between structural and magnetic degrees of freedom.

The principal techniques available to study magnetic properties under high, pressure are: i) susceptibility measurements by the inductance method, ii) neutron diffraction, iii) Mössbauer spectroscopy and iv) inelastic x-ray scattering.

Measurements of susceptibility in large volume pressure cells, where the whole measuring unit is placed inside the pressurized volume, are common but limited in pressure. Measurements in a DAC are more difficult because of the impossibility of placing coils within the pressurized volume. Nevertheless many coil arrangements in DACs exist (Eremets, 1996). Pressures up to 15 GPa (Cornelius *et al.*, 1996) and temperatures of 30mK have been reported (Webb *et al.*, 1976).

Neutron diffraction is very sensitive to the magnetic order (Goncharenko & Mirebeau, 1998), but the determination of the magnetic moment variations remains limited in precision (Link *et al.*, 1998). Although most published work is reported for pressures below 20 GPa, recent studies up to 50 GPa in a DAC at liquid He temperature have been performed (Goncharenko & Mirebeau, 1997).

Mössbauer spectroscopy with nuclear forward scattering of synchrotron radiation is also used to observe pressure-induced magnetic hyperfine interactions (Nasu, 1998). Pressure limits are between 40 and 100 GPa, depending on the radioactive sources, with the exception of 200 GPa for $^{57}$Fe. Synchrotron sources will probably allow measurements at pressures above 100 GPa. The main problem of Mössbauer spectroscopy is the small number of Mössbauer isotopes, which excludes a large number of magnetic materials.

Inelastic x-ray scattering is a very recent and powerful technique enabled by the use of third generation synchrotron radiation sources, which opens the way to studies on any magnetic element in any magnetic order. This is a great advantage compared to Mössbauer spectroscopy and XMCD. Nevertheless this technique is very difficult to handle under pressure.

Compared to these techniques XMCD takes advantage of its element and orbital selectivity, its sensitivity in the detection of very small magnetic moments (0.01 $\mu_B$) and its intrinsic information content on local structural and electronic properties through the XAS signal.

Measuring XMCD at the K-edges of 3d transition metal elements on samples subject to high pressure is a very challenging experiment. In the first place, because the signal is much weaker than at the $L_{II-III}$ edges, since we are not probing the magnetically interesting localized 3d orbitals but the extended 4p bands, the magnetism of which is due merely to the spin-orbit coupling in the final state. In fact, K-edge signals are typically of the order of $10^{-3}$ (or less) in saturation conditions. Secondly, best results for high pressure XAS are obtained in the transmission geometry, where the X-ray beam passes through the highly absorbing diamonds of the high pressure cell. The transmission of the X-ray beam at the Fe-K edge (7100 eV) through a pair of 1.2 mm thick diamonds is lower than 0.4 % and falls below 0.07 % at 6.5 keV. Moreover, to generate circularly polarized X-rays from the linearly polarized X-ray beam provided by the two planar undulators installed on the straight section of ID24, we use a 0.7 mm thick diamond quarter wave plate, that introduces additional absorption at low energies. Last but not least, to reach high pressures of 20-100 GPa, the anvil culets must not exceed diameters of 450 to 150 μm. The gasket hole diameter is approximately 1/3 of this value, and the sample size is therefore limited to about 100 – 50 μm or below.

Despite all these difficulties it is of great interest to be able to perform high pressure XMCD measurements in the low energy hard X-ray range since this represents the only way to have access to this information, in the pressure range covered by the DAC, on 3d transition metals. This energy range also covers a large part of the L-edges of rare earths. Advantages in the use of hard X-rays (K-edges rather than L-edges for 3d metals, L-edges rather than M-edges for rare earths) are also to be attributed to their higher bulk sensitivity.

In the next section we give a short description of the XMCD process at K-edge of 3d transition metal elements and highlight the importance of such measurements at high pressure, and in particular when, in combination with XAS spectra, correlation between structural and magnetic properties is possible. In section 3, we briefly describe the experimental layout for XMCD experiments on beamline ID24. Finally, in section 4, we report a few recent high pressure XMCD results obtained at the Fe-K edge of pure Fe, and on the $Fe_3O_4$ compound.

**2. K-edge XMCD on 3d transition metals**

The interpretation of the XMCD signal at the L edges of transition metals has been clarified by much theoretical work (Thole et al., 1992), (Altarelli, 1993), (Carra et al., 1993), (Ankudinov & Rehr, 1995). Using the sum-rules it is possible to derive the amplitude of both the spin and the orbital magnetic moments from a measurement of the XMCD signal at the $L_2$ and $L_3$ edges. Unfortunately the interpretation of the XMCD signal at the K-edge of transition metals is far from being unanimously agreed upon and no such theory as that mentioned above for the L edges is available. Three different theoretical approaches can be found in the literature: i) a full relativistic multiple-scattering approach (Ebert *et al.*, 1988), (Guo, 1996), (Guo, 1997), (Ankudinov & Rehr, 1997), ii) multiple scattering theory with a non-relativistic description of the electrons (Brouder & Hikam, 1991) and iii) band structure calculations using a tight binding approximation (Igarashi & Hirai, 1994), (Igarashi & Hirai, 1996). But up to now, there is still a lack of a theoretical description able to fully describe the experimental data. It is also for this reason that good quality XMCD data at high pressure will be of fundamental input to help improve theory for the basic understanding of the process.

The reasons underlying the different treatments of K-edge data are related to several factors: the spherical symmetry of the initial state (no spin-orbit coupling in the initial state), the fact that the process does not directly involve the 3d band, responsible for magnetism in 3d metals, and because of the itinerant character of the magnetism in these systems. There is a general agreement that the spin-orbit interaction in the final state (p band) is responsible for the formation of the XMCD signal at the K-edge in 3d transition metals. This interaction is very weak, since it is directly proportional to the orbital kinetic moment $<L_z>$, generally considered to be zero because of the quenching of the crystal field in 3d metals. In reality this orbital moment does give a weak contribution to the magnetic moment. In pure 3d transition metals its values are: 0.09 $μ_B$ for Fe, 0.15 $\mu_B$ for Co and 0.07 $\mu_B$ for Ni. This explains why the XMCD signal at the K-edge is about 100 times weaker than at the $L_{2,3}$-edges in these systems.

To summarize, it is therefore difficult to extract absolute quantitative information from the XMCD signals at the K-edges in transition metals. Nevertheless, the importance in carrying out such experiments at ambient pressure has already been established in the past (Pizzini *et al.*, 1995), (Sakurai et al., 2000). In these works, valuable information on the spin orientation in the p band or on the variation of the element specific magnetic moment with temperature or with its concentration was obtained. The interest in extending this technique to high pressures is fundamental (i.e. as input for theoretical models, where high pressure data is practically non-existent), but can also have important and far reaching implications in geophysics, where these studies could eventually be extended to a larger portion of P-T phase space representative of the interior of many planets including the Earth. Such studies on bulk transition metal compounds (oxides, silicates, etc...) of direct interest to geophysics would then directly lead to a better insight into correlations between structural and magnetic transitions. The extension to low temperatures will offer also very exciting opportunities in the study of strongly correlated systems.

## 3. The energy dispersive XAS beamline ID24 for high pressure XMCD

### 3.1 Introduction

Beamline ID24 is the ESRF XAS beamline with parallel detection of the whole spectrum made possible by energy dispersive highly focusing X-ray optics. The absorption spectrometer is coupled to two low K undulators in a high β section of the ring, through a Kirkpatrick Baez (KB) optical system. The energy range of operation is 5 – 27 keV. The spectral characteristics required for experiments at selected absorption edges are obtained by gap tuning and tapering the undulators. The original and the recently upgraded design of the beamline can be found elsewhere (Hagelstein *et al.*, 1993, Hagelstein *et al.*, 1997, Pascarelli *et al.*, 2004). The choice of an undulator for energy dispersive XAS applications is unusual, but there are clear advantages in using a high brilliance source for high pressure XMCD at low energies. Besides the high flux, the small source size coupled with appropriate defocusing optics yields the required focal spot for high pressure studies, and the low monochromatic divergence allows the optimal exploitation of the properties of quarter wave plates to tune the helicity of incoming photons. In the year 1999, a local feedback system was installed in the ID24 straight section to control the horizontal instabilities of the electron beam (Pascarelli & Plouviez, 2000). This implementation led to a substantial improvement in the stability of the focal spot and in the quality of the XMCD data collected on ID24.

This beamline is now particularly suited for high pressure XMCD measurements using diamond anvil cells. The undulator source, the absence of mechanical movements of the spectrometer during the acquisition of the spectra and the strongly focusing optical scheme of the beamline allow covering the necessary requirements for high pressure XMCD experiments.

    1- A small focal spot. The required focal spot dimensions for applications in the 20-100 GPa region are typically limited to 20x20 $\mu m^2$ full width. The

undulator beam on ID24 undergoes double focusing in both directions and its dimensions, in the 5-9 keV energy range of interest to these studies, are of this order of magnitude.

2- A stable beam. The required signal-to-noise ratio is obtained by averaging a large number of spectra. During the acquisition of a series of spectra to be averaged, the beam has to have excellent spectral and spatial stability. This is possible thanks to the dispersive optics scheme.

3 – Time resolution. Real time visualization of the entire absorption spectrum, intrinsic to the dispersive scheme, is extremely useful to align the anvils in such a way that diffraction from the diamonds is avoided in the energy range of interest. Moreover, very short acquisition times allow to average over a large number of spectra in a reasonable time. Typically, an acquisition averaged over 200 spectra can be carried out in 30 minutes.

4 - High flux. To compensate the high level of absorption in the diamond cell and in the quarter wave plate and to reach the required signal-to-noise level the highest flux possible is needed.

After many years of difficulties, important improvements in the optics (Pascarelli *et al.*, 2004), (Mathon *et al.*, 2004) and beam stability on ID24 (Pascarelli & Plouviez, 2000) lead us recently to obtain good S/N high-pressure XMCD signals at the Fe K edge. This work is the result of years of collaboration between the ID24 team and the Physique des Milieux Condensés laboratory at the University of Paris VI (PMC). Similar studies have been carried out at other synchrotrons at the L-edges of 5d transition metals (Odin et al., 1998), (Odin et al., 1999) as well as at the K-edge of Fe (Ishimatsu et al., 2001) (Ishimatsu et al., 2003).

## 3.2 The optical scheme

Figure 1 illustrates the present optical scheme for XMCD measurements. It consists of a pair of mirrors in a Kirkpatrick-Baez geometry, the Si (111) polychromator, the diamond quarter wave plate (QWP), the third vertically refocusing mirror, the sample environment (the Diamond Anvil Cell, a spectrometer for pressure calibration and the electromagnet) and the position sensitive detector that transforms the energy-direction correlation into an energy-position correlation (Koch *et al.*, 1996). Additional details on the standard optical setup may be found in (Hagelstein *et al.*, 1997), (Pellicer Porres *et al.*, 1998) (San Miguel *et al.*, 1998) (Pascarelli *et al.*, 2004).

At the energies corresponding to the K-edges of 3d transition metals, the full dimensions of the focal spot are below 20 μm in both directions. And the total energy range ΔE diffracted by the polychromator is of the order of 300 – 400 eV, allowing having access to the edge region (XANES) and to the first few EXAFS oscillations (up to a photoelectron wavevector $k \sim 8 - 10$ Å$^{-1}$)

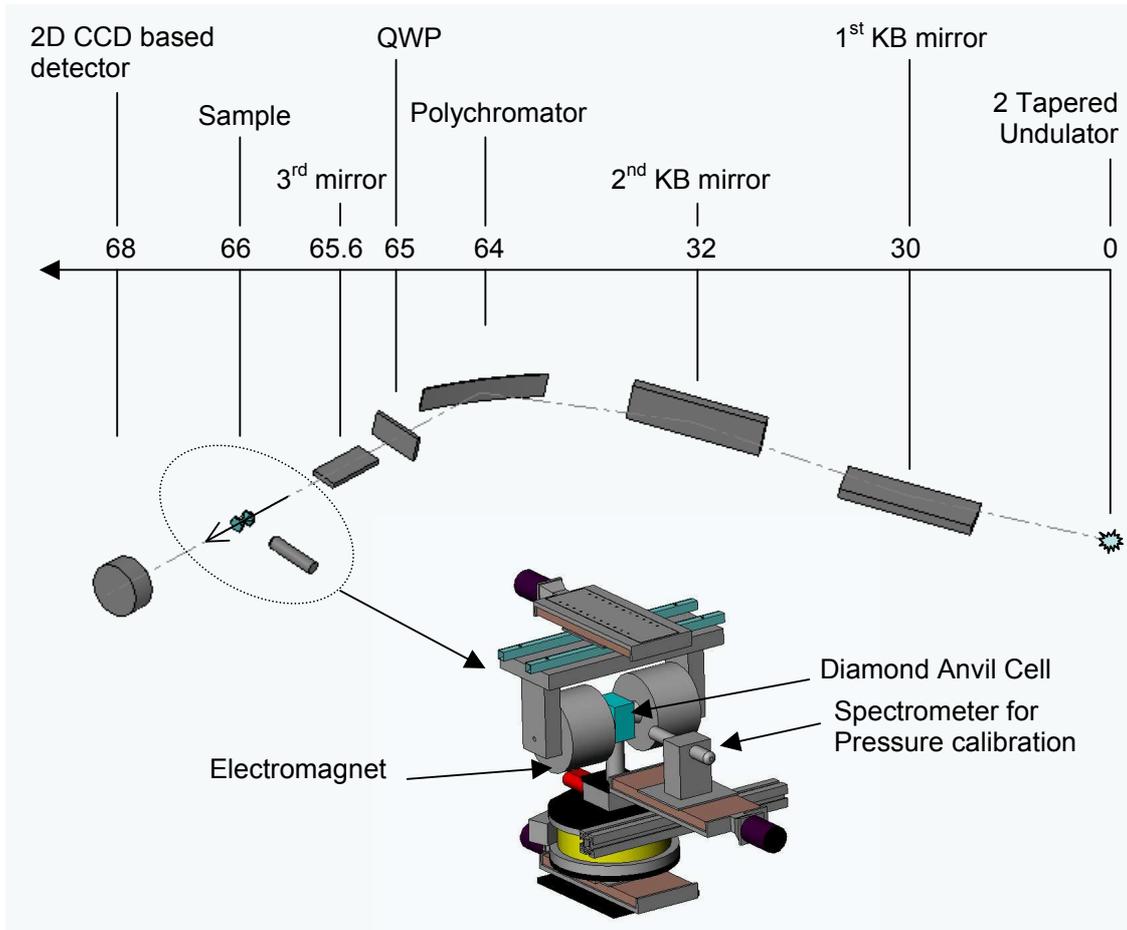

Figure 1 : ID24 optical scheme and sample environment for XMCD at high pressure (not scaled).

The quarter wave plate is located 750 mm downstream the polychromator. It consists, for the energy range of K-edge of the 3d transition metals, in a thin (740 μm) perfect diamond crystal in the Laue geometry. Its birefringence properties close to a Bragg reflection transform the linearly polarized incident X-ray beam into a transmitted (non-deviated) beam with circular polarization very close to unity (Giles *et al.*, 1994). The very small source divergence and the non-dispersivity condition allow circular polarization close to unity to be obtained on the whole polychromatic and divergent fan (Pizzini *et al.*, 1998). XMCD applications on ID24 currently cover a wide energy range, that extends from the L edges of rare earths (down to the $L_3$ edge of Ce at 5.7 keV) up to the L edges of 5d metals (up to the $L_2$ edge of Pt at 13.3 keV). To cover this wide energy range, ID24 disposes of a set of three quarter wave plates (two diamond plates and one Si plate) designed and commissioned on ID24 in collaboration with C. Malgrange of the LMCP of the Paris 6 University.

The third mirror is installed on the polychromatic beam after the QWP, at 0.6 m upstream the focal point. A complete description and the performance of this mirror can be found in (Pascarelli *et al.*, 2004) and (Hignette *et al.*, 2001).

A new sample environment has been specifically developed for the high pressure XMCD applications, in order to concentrate in a compact volume the high pressure cell, the laser spectrometer for the pressure measurement and the magnetic field. The high pressure cell is made of non-magnetic CuBe, manufactured by the Laboratory PMC. Its design derives from the Chervin type membrane cell (Chervin *et al.*, 1995). The HP-DAC is located on a stack of motorization stages that allows its alignment to the beam: vertical Z and horizontal X (transverse to the beam direction) translation stages, rotation around the vertical Z axis and around the horizontal X axis and rotation around the X-ray beam direction. The last three stages are used to remove diffraction peaks originating from the diamonds that fall in the energy range of interest. The use of perforated diamonds has shown to be useful to limit the absorption of the X-ray beam by the diamonds (Dadashev *et al.*, 2001). We have recently tested such diamonds in the 5-7 keV range on D11 (LURE, Orsay) (Itié *et al.* 2003) and on ID24 (San Miguel *et al.*, 2004) up to 30 GPa. Due to the thickness of the HP-DAC (30 mm) the magnetic field obtained on the sample with our classical electromagnetic coil was limited to 0.4 Tesla.

### 3.3 Methodology and performance at ambient pressure

The XMCD signal is defined as follows:

$$XMCD = \mu^L - \mu^R = \ln\frac{I_0^L}{I_1^L} - \ln\frac{I_0^R}{I_1^R}$$

where $\mu^L$ and $\mu^R$ are the X-ray absorption coefficients measured using left and right circular polarization (LCP and RCP) respectively.

If we assume that the incident intensity does not change in time, then $I_0^L = I_0^R$ and

$$XMCD = \mu^L - \mu^R = \ln\frac{I_1^R}{I_1^L}$$

In reality, this assumption is wrong, since the incident intensity decreases with time with electron beam current. By adopting the following sequence for the measurement of the transmitted intensity $I_1$ :

$$I^{0_L}, I^{1_R} I^{2_L}, I^{3_R} \ldots, I^{2n-2_L}, I^{2n-1_R} I^{2n_L}$$

(where n is number of the $I_1^L$, $I_1^R$ pairs of acquisitions) and the following algorithm for the calculation of the XMCD signal:

$$XMCD = \frac{1}{2n}\ln\left(\frac{I^{0_L}\left(I^{2_L}\right)^2 \ldots \left(I^{2n-2_L}\right)^2 I^{2n_L}}{\left(I^{1_R}\right)^2 \left(I^{3_R}\right)^2 \ldots \left(I^{2n-1_R}\right)^2}\right)$$

we effectively suppress linear and exponential drifts of different origins and work exclusively with the transmitted intensities. This is an advantage in energy dispersive XAS, where it is as yet not possible to measure simultaneously the incident and transmitted intensities, although work is being devoted to the development of this acquisition scheme. The XMCD signal can therefore be directly reconstructed by measuring only $I_1^L$ and $I_1^R$ (or $I_1^+$ and $I_1^-$ if the magnetic field is flipped instead of the circular polarization).

In order to reduce systematic errors, we record, at each pressure point, the 4 symmetry cases of the XMCD signal obtained by flipping the helicity of the incoming photons (RCP/LCP) and the magnetic field +/- with the following sequences : RCP field +, RCP field -, LCP field – and LCP field +. The XMCD signals are then calculated independently for the two RCP and LCP cases. Taking the average of these 2 measurements greatly reduces systematic errors and highlights the magnetic origin of the calculated signal.

Figure 2 presents the normalized XMCD and XAS signals obtained in ambient conditions (without the DAC) on a pure Fe foil using a magnetic field H = 0.5 Tesla. The amplitude of the XMCD signal is ~ $3.2 \cdot 10^{-3}$ This value is comparable to the amplitude of $2 \cdot 10^{-3}$ calculated by (Ankudinov & Rehr, 1997) showing that the measurement procedure and the beamline setup are correct for both right and left polarization (in particular the degree of circular polarization obtained with the quarter wave plate is close to unity).

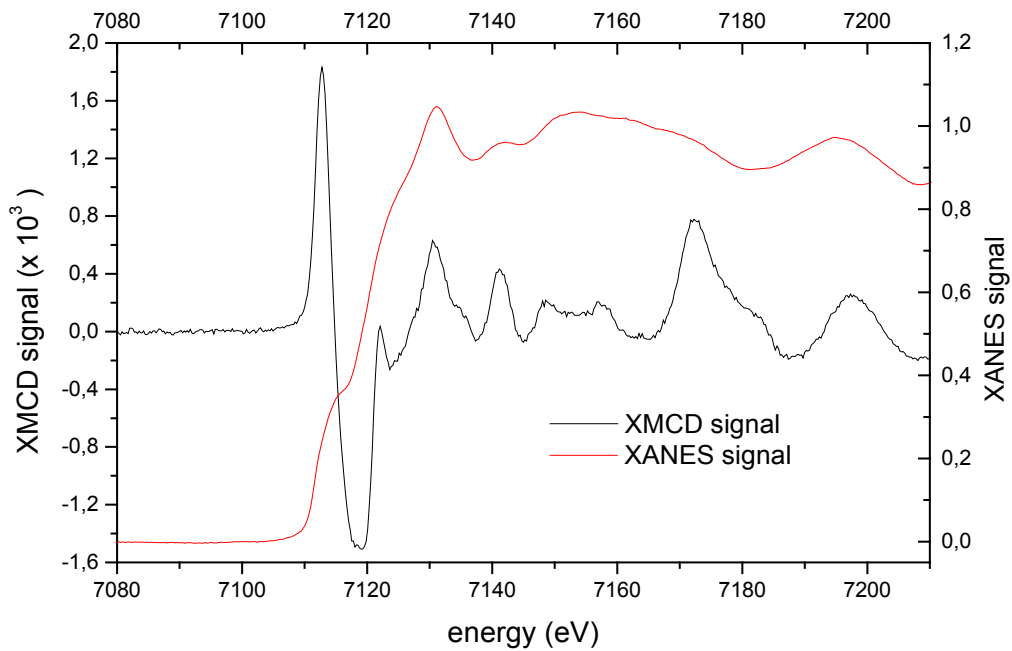

Figure 2 : XAS and XMCD signal recorded on ID24 at the K edge of a pure Fe foil out of the pressure cell.

## 4. Examples at Fe-K edge : the case of pure Fe and $Fe_3O_4$

### 4.1. Pure Fe foil under pressure

Fe exhibits a phase transition from the α ferromagnetic phase to the ε non magnetic phase at around 15 GPa. This transition has been the subject of many structural (mainly XRD) and magnetic studies. Using XAS, it was possible to follow the reorganization of the local structure during the fcc – hcp phase transition (Wang & Ingals, 1998). The two phases were found to coexist in a large pressure domain starting at 13 GPa and spreading over 8 GPa. The evolution of the magnetic moment with pressure has been followed by Mössbauer spectroscopy (Taylor *et al.*, 1991) and by X-ray emission spectroscopy (Rueff *et al.*, 1999). These studies agree that at the end of the transition, the iron bears no magnetic moment. X-ray emission indicates that the Fe magnetic moment follows roughly the bcc phase fraction extracted from a previous XAS study (Wang & Ingals, 1998).

By recording simultaneously XAS and XMCD we can correlate with high precision the changes in the magnetic moment on Fe with an accurate evaluation of the bcc phase fraction during the transition. This gives us the opportunity to conclude whether the magnetism follows exactly the bcc phase stoichiometry during the phase transition. As an example, figure 3 shows XMCD and XANES signals recorded at pressures of 10 and 15.4 GPa. These signals have been recorded on a pure iron foil of 50x50x4 $\mu m^3$ using 400 μm culet diamonds and an Inconel gasket with a 200 μm hole. Silicon oil was used as a pressurizing medium. Pressure was measured using the ruby fluorescence method (Forman *et al.*, 1972). These XMCD profiles are typically obtained using integration times between 1 and 2 hours.

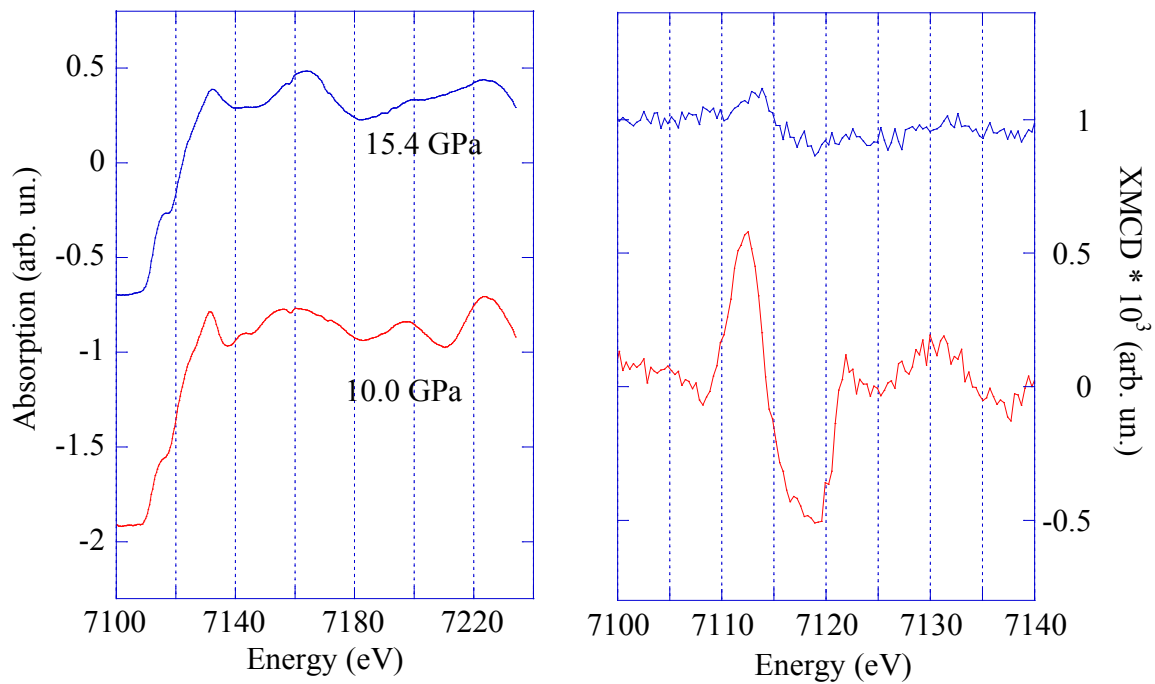

Figure3: Fe-K edge XAS (left panel) and XMCD (right panel) signals of pure Fe foil recorded at 10 and 15.4 GPa. The small glitches in the XANES profile at 7158 and 7171 eV are artifacts
due to small defects of the polychromator crystal.

**4.2. Magnetite under pressure**

A combined XANES-XMCD study has been performed on µm sized magnetite ($Fe_3O_4$) powder between 0 and 30 GPa at room temperature. Below 6 GPa, magnetite crystallizes in the inverse spinel structure in which tetrahedral A sites contain one third of the Fe ions as $Fe^{3+}$, while octahedral B sites contain the remaining Fe ions with equal number of $Fe^{3+}$ and $Fe^{2+}$. Below 860 K magnetite shows ferrimagnetic behavior with magnetic moments on site A aligned antiparallel to magnetic moments on site B (Anisimov *et al.*, 1996). Above 6 GPa the phase transition process is still under discussion (Rozenberg *et al.*, 1996), (Todo *et al.*, 2001).

Figure 5 shows normalized XMCD and XANES signals obtained at 10 GPa. A clear XMCD signal is visible that exhibits two components. The data is currently being analyzed. This data has been obtained using a 400 µm culet diamonds and a 30 µm thick Inconel gasket with a 200 µm diameter hole. Silicone oil was used as pressurization medium. The mass ratio between the powder and the silicone oil is adjusted to obtain a spectrum with a jump approximately equal to 1.

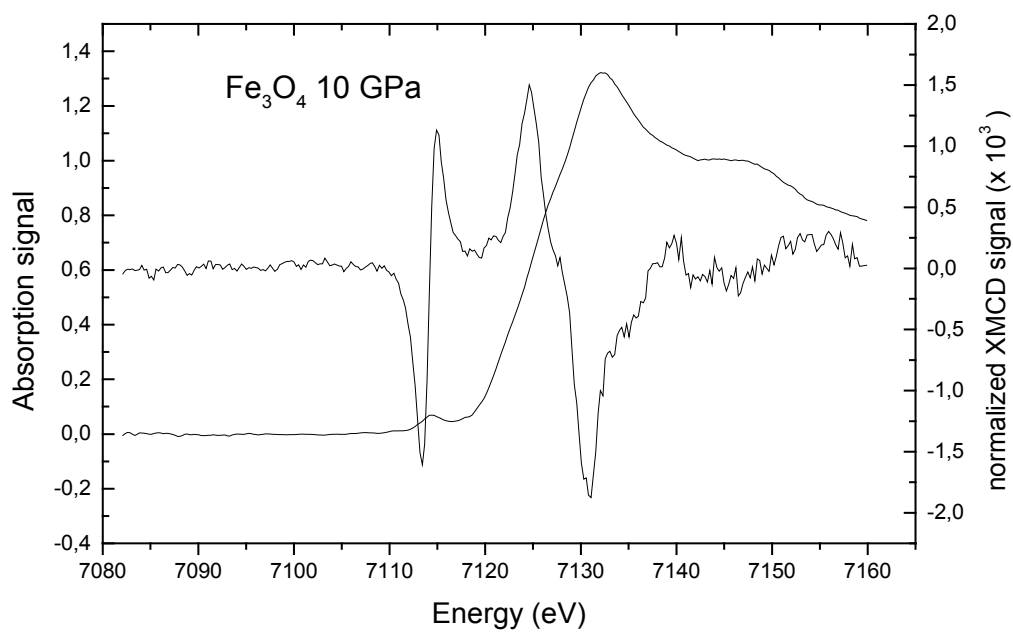

Figure 5 : Fe K edge XMCD and XANES signals measured at 10 GPa on $Fe_3O_4$.

**5. Concluding remarks**

We demonstrate the possibility to record high S/N XMCD and XANES spectra under high pressure at the Fe K edge on the ID24 energy dispersive beamline of the ESRF. XANES and XMCD carried out in the same thermodynamical conditions provide information on the correlation between the local structure and the magnetic properties of the Fe sites. This data can contribute to the basic understanding of the XMCD K-edge process as input for theoretical calculations.

Combined with the development of perforated diamond anvils, these studies provide a new tool to probe the magnetism of 3d transition metals and rare earths under pressure. In the near future, we plan to equip our high pressure experimental setup on ID24 with a He cryostat, to open the field to XMCD measurements under pressure at low temperature.

The authors would like to acknowledge the work of the technical and engineering team of ID24 that makes these experiments possible: M.-C. Dominguez, T. Mairs and F. Perrin. We are also very grateful to the ESRF Crystal Laboratory, in particular to P. Villermet and J.-P. Vassali.